# A variational formulation of the governing equations of ideal quantum fluids.

Roberto Mauri[1,*], Massimiliano Giona[2]

[1] DICI, Department of Civil and Industrial Engineering, Università di Pisa, Largo Lazzarino, 56122 Pisa, Italy.
[2] DICMA, Faculty of Industrial Engineering, La Sapienza, Università di Roma, Via Eudossiana 18, 00184 Rome, Italy.
e-mail:   roberto.mauri@unipi.it   massimiliano.giona@uniroma1.it

**Abstract.**

Applying the least action principle to the motion of an ideal gas, we find Bernoulli's equation where the local velocity is expressed as the gradient of a velocity potential, while the internal energy depends on the interaction among the particles of the gas. Then, assuming that the internal energy is proportional non-locally to the logarithm of the mass density and truncating the resulting sum of density gradients after the second term, we find an additional Bohm's quantum potential term in the internal energy. Therefore, the Bernoulli equation reduces to the Madelung equation, revealing a novel classical description of quantum fluids that does not require to postulate quantum mechanics. Finally, non-locality can be removed by introducing a retarded potential, thus leading to a covariant formulation of the quantum potential and of the equation of motion of an ideal quantum fluid.

## 1. Introduction.

In two recent articles [1,2], we have shown that the internal energy of an ideal gas, assuming a non-local logarithmic dependence from its mass density, is the sum of a classical term and a non-local part where the latter, at leading order, can be expressed as a Bohm potential. Consequently, the continuity and Bernoulli equations that are obtained imposing mass and momentum conservation, at leading order are transformed into the Madelung equation, that is equivalent to a Schrödinger equation. In this work we intend to derive this result using a variational approach, namely the principle of least action.

Consider an ideal gas, composed of point-like particles that do not interact with each other. Since this system does not dissipate, it is natural to derive its governing equations by applying Hamilton's least action principle, as shown by Herivel [3] (see the review articles by Kambe [4-6]), i.e.,

$$\delta A = 0 \qquad A = \int_{t_1}^{t_2} \int_V L\, dt\, d^3\mathbf{x} \qquad L = T - V_s \qquad (1)$$

where the symbol $\delta$ indicates the path variation. Here $A$ is the action, $L$ is the Lagrangian density (energy per unit volume), $T$ and $V_s$ are the kinetic and potential energy density, respectively, with

$$T = \tfrac{1}{2}\rho\mathbf{v}^2 \qquad V_s = \rho(U_{th}+V_e) \tag{2}$$

where $\rho$ is the mass density, $\mathbf{v}$ is the fluid velocity, and we have expressed the potential energy as the sum of an internal energy (or "intrinsic" energy in Herivel's notation) and the external potential energy ($U_{th}$ is the internal energy per unit mass and $V_e$ the external potential energy per unit mass). As mentioned above, we assume that the fluid is homentropic, that is its entropy per unit mass, $S$, is constant and uniform, $S = const$, so that the internal energy $U_{th}$, which in general is a function of $\rho$ and $S$, here depends only on $\rho$. In fact, from simple scaling arguments [7], we have: $U = \frac{kT}{m}\ln\rho$, where $k$ is Boltzmann's constant, $T$ is the temperature and $m$ is the mass of each of the particles that constitute the gas. Note that for homentropic gases the temperature, like the entropy, is uniform and constant, so that, apart from an irrelevant constant, the internal energy $U$ coincides with the free energy, $F=U+TS$ (in fact, in our past work [1,2], we have considered the free energy as the canonical potential).

---
[*] Author to whom any correspondence should be addressed.

Now, the minimization condition, $\delta A = 0$, is subjected to the constrain of mass conservation, that is:

$$\frac{\partial\rho}{\partial t} + \nabla\cdot(\rho\mathbf{v}) = 0. \tag{3}$$

Therefore, Hamilton's equation becomes:

$$\delta\int_{t_1}^{t_2}\int_V\left[\tfrac{1}{2}\rho\mathbf{v}^2 - \rho(U_{th}+V_e) - \phi\left(\frac{\partial\rho}{\partial t} + \nabla\cdot(\rho\mathbf{v})\right)\right]dt\,d^3\mathbf{x} = 0, \tag{4}$$

where $\phi$ is a Lagrange multiplier. Integrating by parts and considering that $\rho$, $\phi$ and $\mathbf{v}$ are fixed at the outer boundaries, we find:

$$\delta\int_{t_1}^{t_2}\int_V L(\mathbf{v},\rho,\phi)\,dt\,d^3\mathbf{x} = 0; \qquad L = \tfrac{1}{2}\rho\mathbf{v}^2 - \rho(U_{th}+V_e) + \rho\left(\frac{\partial\phi}{\partial t} + \mathbf{v}\cdot\nabla\phi\right). \tag{5}$$

Now, as the total energy is conserved and the endpoints are fixed, Hamilton's formulation of the least action principle reduces to Maupertuis's principle:

$$\delta L = \mathbf{p_v}\cdot d\mathbf{v} + p_\rho d\rho + p_\phi d\phi = \sum_q p_q dq = 0, \tag{6}$$

where $p_q = p_q(q,\dot{q},\nabla q)$ satisfies the Euler-Lagrange relations,

$$p_q = \frac{\delta L}{\delta q} = \frac{\partial L}{\partial q} - \frac{\partial}{\partial t}\frac{\partial L}{\partial\dot{q}} - \nabla\cdot\frac{\partial L}{\partial\nabla q} = 0. \tag{7}$$

When $q=\mathbf{v}$, $\rho$, $\phi$ we obtain, respectively:

$$\mathbf{v} = -\nabla\phi, \tag{8}$$

$$\tfrac{1}{2}v^2 - \left(U_{th} + \rho\frac{dU_{th}}{d\rho}\right) - V_e + \left(\frac{\partial\phi}{\partial t} + \mathbf{v}\cdot\nabla\phi\right) = 0, \tag{9}$$

$$\frac{\partial\rho}{\partial t} + \nabla\cdot(\rho\mathbf{v}) = 0. \tag{10}$$

Eq. (8) shows that the velocity field has a potential, equal to the Lagrange multiplier, as one would expect in homentropic fluids.

Defining the thermodynamic enthalpy $H_{th}$ as

$$H_{th} = U_{th} + \frac{p}{\rho}, \quad \text{i.e.,} \quad H_{th} = \tfrac{kT}{m}(\ln\rho + 1), \tag{11}$$

where $p = \rho^2\,\partial U/\partial\rho = \tfrac{kT}{m}\rho$ is the pressure, substituting Eq. (8) into Eq. (9) we obtain:

$$\frac{\partial\phi}{\partial t} = \tfrac{1}{2}(\nabla\phi)^2 + H_{th} + V_e. \tag{12}$$

Note that $\phi$, as well as $H_{th}$ and $V_e$, are all defined within arbitrary constants. Eq. (12) coincides with the Hamilton-Jacobi equation, as shown in Giona and Mauri [2]. Taking the gradient of Eq. (12), we obtain Bernoulli's equation [8]:

$$\frac{d\mathbf{v}}{dt} = \frac{\partial\mathbf{v}}{\partial t} + \mathbf{v}\cdot\nabla\mathbf{v} = \mathbf{F}_S = -\nabla H_{th} - \nabla V_e, \tag{13}$$

with $d/dt$ denoting the material derivative. Naturally the Bernoulli equation, imposing the conservation of energy, is a scalar equation. At steady state, this equation can also be written as:

$$\nabla\left(\tfrac{1}{2}v^2 + U_{th} + \frac{p}{\rho} + V_e\right) = 0. \tag{14}$$

Note that substituting Eq. (8) into Eq. (5) we obtain Fetter-Walecka's Lagrangian density [9],

$$L = \rho\frac{\partial\phi}{\partial t} - \tfrac{1}{2}\rho(\nabla\phi)^2 - \rho(U_{th} + V_e). \tag{15}$$

Applying the Euler-Lagrange Eq. (7) when $q = \rho, \phi$ to Eq. (15) we obtain Eqs. (9) and (10).

Comparing Eqs. (15) and (12) we obtain:

$$L = p, \tag{16}$$

showing that the action to be minimized coincides with the pressure variations (both in time and in space). Using a different route this principle was demonstrated by Seliger and Whitham [10] (see also Eq. (12) in Khalifa and Taha [11].

## 2. The quantum potential

In the previous Section, we have considered that, as $\ln \rho$ and the internal energy $U$ are additive integral of motion, they are proportional to each other. Generalizing this statement, we may assume that this relation is non-local, that is $U$ has the following form:

$$U(\mathbf{r},t) = \frac{kT}{m} \int_{\tau_\infty} u(|\mathbf{r}-\mathbf{r}'|) \ln \rho(\mathbf{r}',t) d^3\mathbf{r}', \tag{17}$$

where $\tau_\infty$ is the total volume, that we assume to be infinite, while $u(x)$ is an interaction kernel between particles located at a distance $x = |\mathbf{r}'-\mathbf{r}|$, with the normalization condition, $\int u(x) d^3\mathbf{x} = 1$. Dropping for convenience the time dependence, expanding $\ln \rho$ in Taylor series,

$$\ln \rho(\mathbf{r}+\mathbf{x}) = \ln \rho(\mathbf{r}) + \mathbf{x} \cdot \nabla \ln \rho(\mathbf{r}) + \tfrac{1}{2} \mathbf{xx} : \nabla\nabla \ln \rho(\mathbf{r}) + \cdots \tag{18}$$

and truncating the series after the second term (there can be no $\nabla \ln\rho$ term, due to the isotropy of the fluid), we find,

$$U(\mathbf{r}) = U_{th}(\mathbf{r}) + \Delta U(\mathbf{r}). \tag{19}$$

Here, the first term on the RHS is the usual classic (i.e., thermodynamic) internal energy (per unit mass) of an ideal gas,

$$U_{th}(\mathbf{r}) = \frac{kT}{m} \ln \rho, \tag{20}$$

while

$$\Delta U = -\tfrac{1}{2} \frac{kT}{m} a^2 \nabla^2 \ln \rho \tag{21}$$

is the non-local part, with,

$$a^2 = -\int_{\tau_\infty} x^2 u(x) d^3\mathbf{x} \tag{22}$$

denoting the square of a characteristic length, $a$. Note the negative sign in Eq. (20), revealing that particles attract each other. Higher-order terms in the expansion (18) can be neglected under quite general conditions, as discussed in Mauri [1].

When the non-local part of the energy is inserted within Hamilton's equation (4), i.e.,

$$\delta \int_{t_1}^{t_2} \int_V \left[ \tfrac{1}{2}\rho \mathbf{v}^2 - \tfrac{kT}{m}\rho\left(\ln\rho - \tfrac{1}{2}a^2\nabla^2\ln\rho\right) - \rho V_e - \phi\left(\frac{\partial\rho}{\partial t} + \nabla\cdot(\rho\mathbf{v})\right)\right] dt\, d^3\mathbf{x} = 0, \quad (23)$$

integrating by parts we find Eq. (5), i.e. $\delta\int_{t_1}^{t_2}\int_V L(\mathbf{v},\rho,\phi)\,dt\,d^3\mathbf{x}=0$, with:

$$L = \tfrac{1}{2}\rho\mathbf{v}^2 - \rho(U_{th}+V_e) + L_Q + \rho\left(\frac{\partial\phi}{\partial t}+\mathbf{v}\cdot\nabla\phi\right), \quad (24)$$

where

$$L_Q = -\tfrac{1}{2}\tfrac{kT}{m}a^2\rho(\nabla\ln\rho)^2 = -\tfrac{1}{2}\tfrac{kT}{m}a^2\rho^{-1}(\nabla\rho)^2. \quad (25)$$

This extra term does not lead to any change within Eqs. (8) and (10), while it introduces an additional term in the Euler-Lagrange Eq. (9), obtaining:

$$\frac{\partial\phi}{\partial t} = \tfrac{1}{2}(\nabla\phi)^2 + H_{th} + U_Q + V_e. \quad (26)$$

where

$$U_Q = \frac{\partial L_Q}{\partial\rho} - \nabla\cdot\frac{\partial L_Q}{\partial\nabla\rho} = -\tfrac{kT}{m}a^2\left[\nabla^2\ln\rho + \tfrac{1}{2}(\nabla\ln\rho)^2\right] = -2\tfrac{kT}{m}a^2\frac{\nabla^2\sqrt{\rho}}{\sqrt{\rho}}. \quad (27)$$

The detailed calculation leading to the last term on the RHS of Eq. (27) can be found in Mauri [1].

As shown in [1,2], when $a$ is the thermal de Broglie wavelength,

$$a = \frac{\hbar}{\sqrt{4mkT}}, \quad (28)$$

then $U_Q$ reduces to Bohm's quantum potential [12-14], i.e.,

$$U_Q = -\frac{\hbar^2}{2}\frac{\nabla^2\sqrt{\rho}}{\sqrt{\rho}}. \quad (29)$$

A beautiful discussion about the consequences of Bohm's theory of motion can be found in the review paper by Holland [15]. Taking the gradient of Eq. (26), we obtain the following modified Bernoulli's equation:

$$\frac{d\mathbf{v}}{dt} = \frac{\partial\mathbf{v}}{\partial t} + \mathbf{v}\cdot\nabla\mathbf{v} = \mathbf{F}_S = -\nabla(H_{th}+U_Q+V_e), \quad (30)$$

This is an equation of energy and momentum conservation (remember that the velocity is the gradient of a scalar, so that this is a scalar equation), where the force per unit mass, $\mathbf{F}_S$, is the gradient sum of three energies, the thermodynamic enthalpy $H_{th}$, the external force potential, $V_e$, and a quantum, so-called Bohm potential, $U_Q$. Together with the continuity Eq. (3), Eq. (26) is the Madelung equation [16], showing that the complex field

$$\psi(\mathbf{r},t) = \sqrt{\rho} e^{i\phi/\hbar}, \tag{31}$$

with the scalar fields $\rho(\mathbf{r},t)$ and $\phi(\mathbf{r},t)$ determined from Eqs. (3) and (26), identically satisfies Schrödinger's equation. Using the identity $\mathbf{v} \cdot \nabla \mathbf{v} = \nabla\left(\tfrac{1}{2}mv^2\right)$, the Madelung equation (26) can be rewritten as:

$$m\frac{\partial \mathbf{v}}{\partial t} = -\nabla E = -\nabla\left[\tfrac{1}{2}mv^2 + H_{th} + U_Q + V_e\right], \tag{32}$$

where the RHS denotes the gradient of the total energy, $E$, equal to the sum of kinetic, internal, external, and quantum potential energy. Eqs. (30) and (32) are the superfluid hydrodynamical equations [17].

## 3. Higher order terms

As shown in [1], using the full expansion (18), we obtain a generalization of Eqs. (19)-(22) with

$$U(\mathbf{r}) = \tfrac{kT}{m} \sum_{n=0}^{\infty} (-1)^n a^{2n} \frac{c_{2n}}{(2n)!} \nabla^{2n} \ln \rho \tag{33}$$

with $c_0 = 1$ and $c_2 = 1$, leading to the first two terms of the expansion that we obtained in the previous Section, including the definition (22) of the characteristic length $a$, while:

$$c_{2n} = (-1)^n \frac{1}{a^{2n}} \int_{\tau_\infty} x^{2n} u(x) d^3\mathbf{x}; \quad n = 2, 3, \cdots \tag{34}$$

Now, substituting Eq. (33) into Eq. (6), with the Euler-Lagrange equation

$$p_q = \frac{\delta L}{\delta q} = \frac{\partial L}{\partial q} - \sum_{n=1}^{\infty} \left(\frac{\partial}{\partial t^n} \frac{\partial L}{\partial \dot{q}^n} + \nabla^n \cdot \frac{\partial L}{\partial \nabla^n q}\right) = 0, \tag{35}$$

we obtain a generalization of the Madelung equation, that is Eq. (27) with [1]:

$$U_Q(\mathbf{r}) = \tfrac{kT}{m} \sum_{n=1}^{\infty} (-1)^n a^{2n} \frac{c_{2n}}{(2n)!} \left(\nabla^{2n} \ln \rho + \frac{1}{\rho} \nabla^{2n} \rho\right) \tag{36}$$

## 4. Retarded potential

The non-local constitutive equation (17) implies that a density change in **r'** determines an instantaneous variation of the internal energy at **r**, with an infinite propagation velocity. To correct this obviously unrealistic assumption, Eq. (17) is replaced with the following expression:

$$U(\mathbf{r},t) = \frac{kT}{m} \int_{V_\infty} \int u(|\mathbf{r}-\mathbf{r}'|) \ln \rho(\mathbf{r}',t') d^3\mathbf{r}' + V(\mathbf{r},t), \tag{37}$$

where

$$t' = t - \frac{|\mathbf{r}-\mathbf{r}'|}{c} \tag{38}$$

is a retarded time and $c$ denotes the propagation velocity.

Expanding $\ln \rho$ in Taylor series,

$$\ln \rho\left(\mathbf{r}+\mathbf{x}, t+\frac{x}{c}\right) = \ln \rho(\mathbf{r},t) + \mathbf{x}\cdot\nabla\ln\rho(\mathbf{r}) + \frac{x}{c}\frac{\partial}{\partial t}\ln\rho(\mathbf{r}) + \tfrac{1}{2}\mathbf{xx}:\nabla\nabla\ln\rho(\mathbf{r}) + \tfrac{1}{2}\frac{x^2}{c^2}\frac{\partial^2}{\partial t^2}\ln\rho(\mathbf{r}) + \cdots \tag{39}$$

and truncating the series after the quadratic terms, we find Eq. (17), i.e., $U = U_{th} + \Delta U$, where

$$\Delta U = \tfrac{1}{2}\frac{kT}{m}a^2\left(\frac{1}{c^2}\frac{\partial^2}{\partial t^2} - \nabla^2\right)\ln\rho = \tfrac{1}{2}\frac{kT}{m}a^2 \Box \ln\rho, \tag{40}$$

with

$$\Box = \frac{1}{c^2}\frac{\partial^2}{\partial t^2} - \nabla^2 \tag{41}$$

denoting the D'Alambertian wave operator. Eq. (40) is the Lorentz-invariant form of the non-local energy (21). Then, proceeding as in the previous Section, we find Eq. (26), where the Bohm potential (29) becomes:

$$U_Q = \frac{\hbar^2}{2}\frac{\Box\sqrt{\rho}}{\sqrt{\rho}}. \tag{26}$$

As shown in Mauri, Giona [2], this allows to write the equation of motion of a quantum ideal gas in covariant form.